# Derivation of Novel Isotropic Metamaterials at the Hashin-Shtrikman Stiffness Limit Enabled by Physics Informed Neural Networks

Manish Kumar Singh[1], Dominic Kang Jueh Lim[1,2], Anson Wen Han Cheong[1,2], Adams Wai Kin Kong[3], Chang Quan Lai[1,2,4*],

[1]*School of Mechanical and Aerospace Engineering, Nanyang Technological University, 50 Nanyang Ave, Singapore, 639798, Singapore*

[2]*Singapore Centre for 3D Printing, 63 Nanyang Drive, Singapore 636922, Singapore*

[3]*College of Computing and Data Science, Nanyang Technological University, 50 Nanyang Ave, Singapore, 639798, Singapore*

[4]*School of Materials Science & Engineering, Nanyang Technological University, 50 Nanyang Ave, Singapore, 639798, Singapore*

[*] cqlai@ntu.edu.sg

**Keywords:** Plate Lattice; Metamaterial; Hashin-Shtrikman; Machine Learning; Physics-Informed Neural Networks; LAPIS

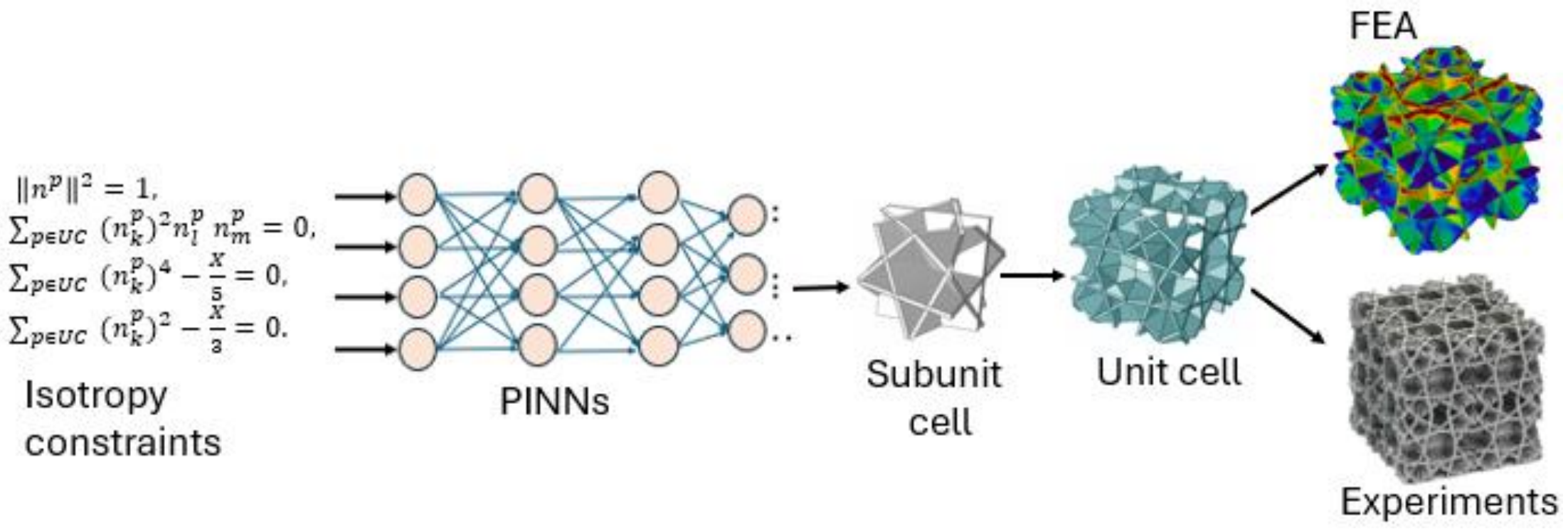

## Abstract

Current methods of deriving ultrastiff isotropic metamaterials require heuristics or topology optimization, which are highly inefficient and limited by human input. To circumvent this, Physics-Informed Neural Networks (PINNs) are applied to analytical equations to systematically solve for plate orientations in closed cell lattices. Following a standardized procedure, the plate orientations were modelled and assembled into geometries resembling quasi-random foams for validation via finite element simulations. Closed cell 6-plate geometries were also additively manufactured using LAPIS and experimentally characterized. The results indicate that a minimum of 6 plates is required to constitute an ultrastiff isotropic lattice, 7 plates would return the established plate Cubic Octet (a.k.a. *p*SC-*p*FCC) structure while 8- to 10- plate geometries did not show significant differences in properties. Despite the introduction of some plate bending due to sub-optimal plate placement strategy, relative stiffness of the PINNs lattices adheres closely to the Hashin–Shtrikman upper bound – 90% of Hashin-Shtrikman upper bound at $\rho_R$ = 0.4, and the lattices maintain near-isotropic behaviour, achieving a Zener ratio of 0.95 at $\rho_R$ = 0.1. Benchmarking results show that specific stiffness of PINNs lattices is higher than the majority of isotropic architectures reported in the literature.

# 1 Introduction

There is a perpetual demand for materials with ever higher specific stiffness as they can directly enhance the fuel efficiency of automobiles and aerospace vehicles[1–3], improve the performance of sports equipment such as tennis rackets and golf clubs[4], as well as increase the efficiency of equipment parts engaged in high-speed automation and production. To this end, modern technologies have often turned to lightweight composites reinforced with carbon fibers[5–7], glass fibers[5,8,9], graphene[10–12] and a variety of micro-[13–15] and nano-fillers[16,17]. To achieve greater weight savings, materials can even be composited with empty spaces to form solid foams[18,19] and lattice metamaterials[20–25] that can attain relative densities as low as ~ $10^{-4}$ [26,27].

Extensive research into the microstructure-mechanics relationships of solid foams[28,29] have shown that their microstructures typically favor beam bending deformation, which limits their stiffness[28]. To address this shortcoming, Ashby *et al.* introduced a new class of truss designs[30,21] with strategically positioned beams that reduce bending and steer the deformation towards beam stretching/ compression instead. Because stretching requires more stress than bending to realize the same strain, these stretch-dominated lattices were able to display a better rigidity than foams of the same density.

The enhanced stiffness, however, came at the cost of isotropy – while foams exhibit the same stiffness in all directions, metamaterial lattices have an orientation-dependent elastic response. This structural anisotropy can (i) significantly complicate design and mechanical analysis of parts in a system, (ii) compromise structural integrity under multi-axial loading and (iii) provide less protection when safety gears (*e.g.* helmets, armor) are impacted at certain angles. Due to these limitations, researchers have long sought metamaterial designs that exhibit isotropy, as well as the loading efficiency of stretch-dominated geometries.

Some of the first isotropic lattices reported were achieved through combinations of Simple Cubic (SC) with Body-Centered Cubic (BCC)[31,32] and/ or Face-Centered Cubic (FCC) trusses[33,34]. The specific relative modulus, $\bar{E}/(E_s\rho_r)$, of these isotropic beam-based lattices, however, were found to be limited to ≤ 0.17 at small relative densities (*i.e.*, $\rho_r \rightarrow 0$)[32,34]. which is far short of the maximum value of ~ 0.5 theorized for isotropic structures, based on Hashin-Shtrikman's seminal analysis[35,36].

As a result of this poor tradeoff between isotropy and specific stiffness for truss designs, recent years have seen a shift of interest from beam-based lattices to plate-based lattices[36–39]. These new lattices often utilize a highly interconnected network of plates to induce stretch-dominated deformation and maximize stiffness, in sharp contrast to the more established Triply Periodic Minimal Surfaces (TPMS)[40–45] and Shellular structures[20,46], which consist of curvaceous membranes that are more susceptible to bending-dominated deformation[47]. Because 2D plates can resist tensile/ compressive loading from all directions within a plane, as opposed to 1D beams that can only do so in a single axis, they offer a higher loading efficiency for a given mass. Consequently, plate lattice metamaterials were able to exhibit specific relative moduli closer to the Hashin-Shtrikman (H-S) upper bound than beam lattices. For instance, by substituting beams for plates in the classic Octet Truss design, the specific stiffness can be increased by more than 3× for a given relative density, although the resultant plate lattice remains anisotropic[36,48].

Taking cues from isotropic truss designs, Berger *et. al.*[36] and Tanconge-Dejean *et. al.*[37] first showed that isotropic plate lattices can be similarly obtained through a combination of SC, BCC[32] and FCC designs in the volume ratio of $x$: 0.4–0.25$x$: 0.6–0.75$x$, where $0 \leq x \leq 0.8$. Unlike truss geometries, however, these isotropic plate lattices were predicted to exhibit stiffnesses at the H-S upper bound. Because the isotropic plate lattices were closed cell structures, early experimental efforts focused on perforated versions of these designs – the added holes were necessary to release uncured resins or unsintered powder from the inner voids after lattice fabrication. Nevertheless, the initial results were encouraging as the lattice stiffnesses were indeed close to the H-S upper limit, as expected from analytical computations[37,49]. These findings had recently been further reinforced, when unadulterated isotropic plate lattices, fabricated using a novel additive manufacturing technique[50]. LAPIS, were tested rigorously for both isotropy and adherence to the H-S upper bound[51].

Since then, more of such lattices have been discovered[38,52–55], albeit at a slow rate. The main challenge lies with the need to first formulate a candidate architecture by hybridizing designs from a limited pool of well-studied geometries. This architecture is then subjected to parametric[38,52,56] or analytical optimization[37,57] to ascertain whether they exhibit both isotropy and high specific stiffnesses. These heuristic efforts are time consuming, resource-intensive and highly dependent on the creativity and experience of individual researchers. Furthermore, the need to combine 2 or more known geometries to produce new isotropic designs will only lead to structures that are increasingly complex and impractical to manufacture.

To circumvent these difficulties, we propose a highly efficient first-principles analytical approach for deriving isotropic plate lattices with stiffnesses at the H-S upper bound, without the need to presuppose any initial geometry for optimization. To help solve the intractable nonlinear algebraic system of constitutive equations, physics informed neural networks (PINNs) were employed. The derived designs were then validated using finite element simulations, as well as experiments on additively manufactured stainless steel samples. The elastic responses obtained were also benchmarked against that of other isotropic structures reported in the literature.

# 2 Materials and Methods

## 2.1 Finite Element Simulations

To validate the analytical solution, Finite Element Analysis (FEA) was conducted using ABAQUS Standard 2025. The effective relative Young's modulus ($\bar{E}/E_s$), shear modulus ($\bar{G}/G_s$), and Poisson's ratio ($\bar{\nu}$) of the plate lattice unit cell were evaluated in the linear elastic regime under periodic boundary conditions, using the EasyPBC plugin tool[58]. This plugin facilitates the application of periodic boundary conditions (PBCs) and extraction of homogenized elastic properties of a representative volume element (RVE) of a composite material (Supplementary Material Figure S1). The computational domain was discretized using 10-node quadratic tetrahedral elements (C3D10M), and a convergence study with respect to mesh density was conducted to identify the optimal mesh size for the simulations (Supplementary Material Figure S2).

## 2.2 Metamaterial Lattice Fabrication

Because the plate lattice designs involve fully enclosed air spaces, Laser Pulsed Integration of Sheets (LAPIS)[50], was employed to fabricate the lattices, as it is currently the only manufacturing technique that has produced true closed cell architected structures, without the need to introduce release holes for draining of precursor materials[51]. The fabrication procedure had been reported previously[51]. Briefly, it begins with the welding of 0.05mm thick SS304L metal foil onto a base substrate using an infrared (IR) laser (wavelength = 1094 nm). This is followed by laser cutting of the outline, according to that layer's input design. Excess material is removed via vacuum suction to prevent entrapment of material within the enclosed cavities. A laser cleaning step, together with mechanical deburring, is then conducted to remove surface oxides and surface asperities generated during processing. These steps are repeated until the lattice structures are fully fabricated.

To remove the anisotropy introduced to the constituent material by the LAPIS process, the samples were heat treated at 1200°C for 2 hours to induce recrystallization[51]. Taking into consideration that the minimum manufacturable plate thickness was 0.20 mm and the maximum fabrication height was 15 mm in the LAPIS setup, the optimal relative density of the stainless steel plate lattices selected for printing was 0.45 (Fig. 3). To investigate the isotropy of the lattices, the structures were printed in <100> and <110> orientations.

## 2.3 Micro-CT Characterization

Zeiss XRadia Versa 620 equipped with a 0.4× objective lens was used to perform micro-CT scanning of the fabricated closed cell lattices. The sample was scanned with 25W of power, at a voltage of 160kV, with an exposure time of 1s per projection. An isotropic voxel size of 15.5μm was used, and reconstruction was conducted using Dragonfly 3D World ZEISS software.

## 2.4 Mechanical Testing

Mechanical testing was performed with Shimadzu Autograph AGS-X equipped with a 50 kN load cell, using a quasi-static strain rate of 0.001 $s^{-1}$. The elastic modulus of the microlattices was determined by tracking the change in vertical distance between their top and bottom surfaces using the open-source Tracker software.

# 3 Derivation of Isotropic Plate Lattice Designs

## 3.1 Constitutive Equations

Consider a metamaterial lattice composed only of straight plates constituted from a homogenous, isotropic material. In the limit of small relative density, where the intersecting volume of the plates is small, the relative density, $\rho_R$, of the metamaterial lattice can be expressed as,

$$\rho_R = \frac{V_p}{V_{UC}} = \frac{\rho_{UC}}{\rho_p} = \sum_{i \in UC} v^i \tag{1}$$

where $V_p$ refers to the total volume of all the plates in the unit cell, $V_{UC}$ refers to the nominal volume occupied by the unit cell, $\rho_p$ refers to the mass density of the plate material, $\rho_{UC}$ refers to the nominal mass density of the unit cell and $v^i$ is the volume fraction of each individual plate – the superscript *i* is a reference number for each individual plate in the unit cell. We will assume that $v^i$ is the same for all the plates, which implies that the thickness of individual plates will be different if the plates have different areas.

The orientation of each plate, $i$, in the lattice can be defined by its unit normal, $\boldsymbol{n}^i = \{n_1^i, n_2^i, n_3^i\}^T$ and follows

$$(n_1^i)^2 + (n_2^i)^2 + (n_3^i)^2 = 1 \tag{2}$$

Previous analysis have shown that, for small, elastic, in-plane deformation of the plates under plane-stress state, the metamaterial lattice will exhibit an isotropic elastic response if [37]

$$\sum_{i \in UC} v^i (n_k^i)^2 n_l^i \, n_m^i = 0, \; for\; k, l, m = 1,2,3 \; with \; l \neq m, \tag{3}$$

$$\sum_{i \in UC} v^i (n_k^i)^4 = \frac{\rho_R}{5}, for\; k = 1,2,3. \tag{4}$$

$$\sum_{i \in UC} v^i (n_k^i)^2 = \frac{\rho_R}{3}, for\; k = 1,2,3. \tag{5}$$

Having satisfied Eq. (3) – (5), the effective relative elastic modulus ($\bar{E}/E_s$), relative shear modulus ($\bar{G}/E_s$) and Poisson's ratio ($\bar{\nu}$) of the isotropic metamaterial lattice will then be[35–37]

$$\frac{\bar{E}}{E_s} = \frac{-2\rho_R(-7+5\nu_s)}{(15\nu_s{}^2\rho_R+2\nu_s\rho_R-15\nu_s{}^2-13\rho_R-12\nu_s+27)}, \quad (6)$$

$$\frac{\bar{G}}{E_s} = \frac{-\rho_R(-7+5\nu_s)}{2(10\nu_s\rho_R-8\rho_R-15\nu_s+15)(1+\nu_s)}, \quad (7)$$

$$\bar{\nu} = \frac{5\rho_R\nu_s{}^2+2\rho_R\nu_s-15\nu_s{}^2-3\rho_R+12\nu_s+3}{15\rho_R\nu_s{}^2+2\rho_R\nu_s-15\nu_s{}^2-13\rho_R-12\nu_s+27}, \quad (8)$$

where $\bar{E}$, $\bar{G}$, and $\bar{\nu}$ refer to the effective stiffness, shear modulus and Poisson's ratio of the lattice, while $E_s$ and $\nu_s$ refer to the Young's modulus and Poisson's ratio of the constituent material. In the limit of small relative density ($\rho_R \rightarrow 0$), Eq. (6) – (8) can be simplified to[37,57]

$$\lim_{\rho_R\to 0}\frac{\bar{E}}{E_s} = \frac{-2\rho_R(-7+5\nu_s)}{(-15\nu_s{}^2-12\nu_s+27)}, \quad (9)$$

$$\lim_{\rho_R\to 0}\frac{\bar{G}}{E_s} = \frac{-\rho_R(-7+5\nu_s)}{30(1-\nu_s{}^2)}, \quad (10)$$

$$\lim_{\rho_R\to 0}\bar{\nu} = \frac{1+5\nu_s}{9+5\nu_s}. \quad (11)$$

where Eq. (9) is essentially the H-S upper bound for elastic modulus of isotropic structures. In addition, it is worth noting from Eq. (9) – (11), that $\bar{E}/E_s$ and $\bar{G}/E_s$ are linearly proportional to $\rho_R$, while $\bar{\nu}$ is independent of $\rho_R$, when the relative density is small ($\rho_R \rightarrow 0$).

To date, the analytical method for deriving isotropic plate lattices from the above equations is to first assume a hybrid design of 2 or more different unit cells with known plate orientations, $\boldsymbol{n}^i$ [37]. This would then leave the plate thickness of each unit cell geometry as the only parameters that can be varied to satisfy Equations (1), (3) – (5). The discovery of novel isotropic geometries using this approach can be rather slow and challenging, as it requires researchers to guess the geometry first, when a large proportion of the guesses will yield no solutions. Moreover, the pool of well-studied structures that researchers can draw from to formulate these guesses can be rather limited, especially since they must be easy enough to be visualized and modelled by humans.

Therefore, for the present investigation, we elected not to presume any initial geometries and assigned the plate orientations, $\boldsymbol{n}^i$, rather than plate thickness ratio, as the key variables to be solved in Eq. (2) – (5), for a given number of plates, $X$, and relative density, $\rho_R$, of the unit cell. Note that $X$ only includes plates with unique orientations *i.e.*, 2 or more parallel plates only count as 1. This approach would generate $3X$ unknowns and $X + 15$ equations, which is challenging to solve through traditional means. For this reason, Physics-Informed Neural Networks (PINNs) will be employed as the residual-minimization framework, as it had previously been shown to be highly effective for solving differential equations[59–61], and is equally applicable to our present system of nonlinear algebraic constraints. For clarity, the geometries derived this way will be referred to as *Xp* PINNs lattices, where $X$ refers to the number of plates and $p$ is abbreviation for 'plates'.

### 3.2 Physics-Informed Neural Networks

To implement the physics-informed neural networks (PINNs), a fixed input vector of dimension $3X$ was passed through a fully connected neural network comprising four hidden layers with 64 neurons each, hyperbolic tangent (tanh) activation functions and a learning rate of 0.001, producing a $3X$-dimensional output (Figure 1). This output was reshaped into $X$ 3D vectors, which were then used to evaluate a set of $X + 15$ nonlinear algebraic constraints (Eq. (2)-(5)). The loss function was formulated as the mean squared error (MSE) of these constraint residuals, and through the use of backpropagation and the Adam optimizer, the neural network iteratively minimized the loss, resulting in solutions that satisfied all the imposed constraints in Eq. (2) – (5) without requiring labelled data. PINNs were applied in the present work scope to solve for architectures where $X \leq 10$, although the algorithms can be generalized for any number of plates. A representative Python code for $X = 8$ can be found in the Supplementary Material.

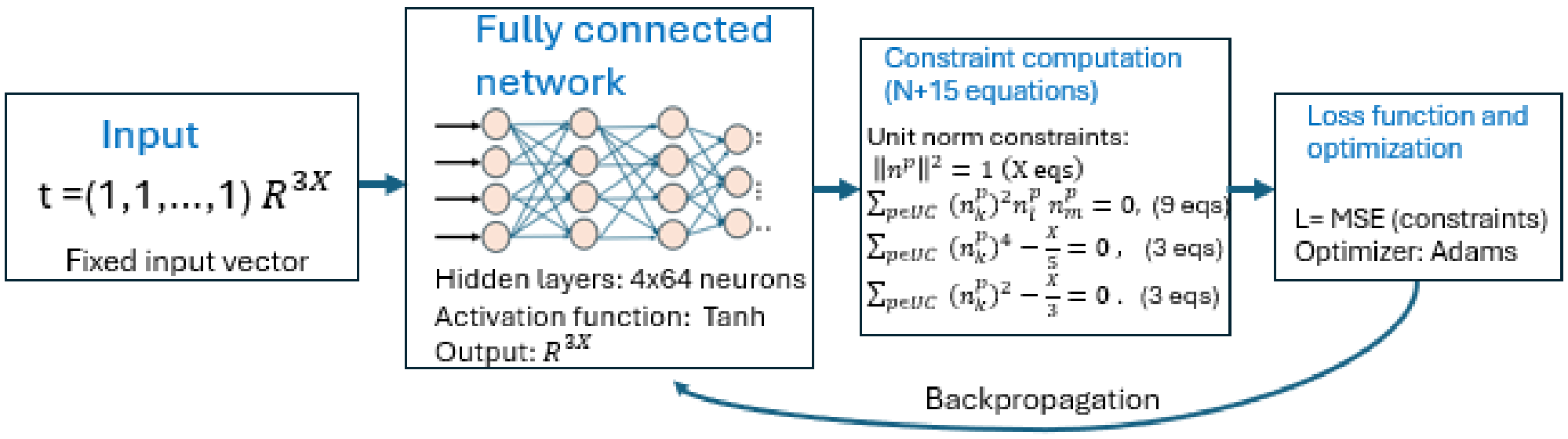

**Figure 1:** Schematic of the Physics informed neural networks (PINNs) framework used in this study**.** A fixed input vector was passed through a fully connected neural network to predict a $3X$-dimensional output. The output was reshaped and used to compute $X + 15$ constraint residuals, which formed the loss function. The network parameters were optimized using backpropagation with the Adams optimizer.

# 4 Results

## 4.1 PINNs Solutions

The application of PINNs to solve for isotropic geometries with $1 \leq X \leq 5$ yielded no results, as the calculations could not converge *i.e.*, the mean squared error (MSE) loss function of all the governing equations (Eq. (2) – (5)) remained elevated ($>10^{-3}$) (Supplementary Material Figure S3). For $X \leq 2$, this is expected, since there are insufficient plates to even satisfy 3D cubic symmetry, while the only structure that is cubic symmetric for $X = 3$, plate Simple Cubic (*pSC*), is known to be anisotropic[36].

For structures consisting of 6 or more unique plates, the MSE loss decreased steadily to a range of $10^{-15}$ to $10^{-4}$ by 30,000 epochs (Supplementary Material Figure S3), indicating that there exist valid plate lattice geometries that could satisfy the isotropy conditions to within a small margin of error. The plate normal vectors solved by PINNs for $6 \leq X \leq 10$ are

$$\boldsymbol{6p} = \begin{Bmatrix} -0.747 \\ 0.416 \\ 0.518 \end{Bmatrix}, \begin{Bmatrix} -0.461 \\ -0.584 \\ 0.668 \end{Bmatrix}, \begin{Bmatrix} -0.110 \\ -0.962 \\ -0.249 \end{Bmatrix}, \begin{Bmatrix} -0.179 \\ -0.195 \\ -0.964 \end{Bmatrix}, \begin{Bmatrix} -0.926 \\ -0.299 \\ -0.232 \end{Bmatrix}, \begin{Bmatrix} -0.572 \\ 0.657 \\ -0.490 \end{Bmatrix}, \tag{12}$$

$$\boldsymbol{7p} = \begin{Bmatrix} -0.984 \\ 0.024 \\ -0.143 \end{Bmatrix}, \begin{Bmatrix} 0.533 \\ -0.832 \\ 0.157 \end{Bmatrix}, \begin{Bmatrix} -0.118 \\ -0.645 \\ 0.754 \end{Bmatrix}, \begin{Bmatrix} 0.694 \\ -0.081 \\ -0.714 \end{Bmatrix}, \begin{Bmatrix} -0.077 \\ 0.758 \\ 0.648 \end{Bmatrix}, \begin{Bmatrix} -0.446 \\ -0.0615 \\ -0.893 \end{Bmatrix}, \begin{Bmatrix} -0.594 \\ -0.804 \\ -0.019 \end{Bmatrix}, \tag{13}$$

$$\boldsymbol{8p} = \begin{Bmatrix} -0.593 \\ 0.807 \\ 0 \end{Bmatrix}, \begin{Bmatrix} 0.917 \\ 0.118 \\ 0.371 \end{Bmatrix}, \begin{Bmatrix} 0.419 \\ -0.388 \\ 0.824 \end{Bmatrix}, \begin{Bmatrix} -0.291 \\ -0.602 \\ -0.744 \end{Bmatrix}, \begin{Bmatrix} -0.917 \\ -0.118 \\ 0.370 \end{Bmatrix}, \begin{Bmatrix} -0.259 \\ -0.966 \\ 0 \end{Bmatrix}, \begin{Bmatrix} 0.419 \\ -0.388 \\ -0.824 \end{Bmatrix}, \begin{Bmatrix} -0.291 \\ -0.602 \\ 0.744 \end{Bmatrix}, \tag{14}$$

$$\boldsymbol{9p} = \begin{Bmatrix} -0.295 \\ 0.888 \\ -0.352 \end{Bmatrix}, \begin{Bmatrix} -0.364 \\ 0.156 \\ 0.918 \end{Bmatrix}, \begin{Bmatrix} 0.977 \\ -0.035 \\ 0.210 \end{Bmatrix}, \begin{Bmatrix} 0.219 \\ -0.444 \\ 0.869 \end{Bmatrix}, \begin{Bmatrix} 0.733 \\ 0.404 \\ -0.547 \end{Bmatrix}, \begin{Bmatrix} -0.778 \\ 0.580 \\ 0.241 \end{Bmatrix}, \begin{Bmatrix} -0.519 \\ -0.854 \\ 0.041 \end{Bmatrix}, \begin{Bmatrix} -0.010 \\ 0.823 \\ 0.567 \end{Bmatrix}, \begin{Bmatrix} -0.605 \\ -0.284 \\ -0.744 \end{Bmatrix}, \tag{15}$$

$$\boldsymbol{10p} = \begin{Bmatrix} 0.568 \\ 0.413 \\ -0.712 \end{Bmatrix}, \begin{Bmatrix} -0.959 \\ -0.072 \\ 0.271 \end{Bmatrix}, \begin{Bmatrix} -0.521 \\ 0.791 \\ 0.321 \end{Bmatrix}, \begin{Bmatrix} -0.499 \\ -0.801 \\ -0.330 \end{Bmatrix}, \begin{Bmatrix} 0.363 \\ 0.289 \\ 0.886 \end{Bmatrix}, \begin{Bmatrix} -0.149 \\ 0.448 \\ -0.881 \end{Bmatrix}, \begin{Bmatrix} -0.929 \\ -0.024 \\ -0.369 \end{Bmatrix}, \begin{Bmatrix} 0.245 \\ 0.953 \\ -0.176 \end{Bmatrix}, \begin{Bmatrix} 0.369 \\ -0.432 \\ -0.823 \end{Bmatrix}, \begin{Bmatrix} -0.595 \\ 0.715 \\ -0.367 \end{Bmatrix}. \tag{16}$$

From Eq. (12) – (16), the angle between faces in the lattices (*i.e.*, dihedral angles), $A$, can be computed using

$$\cos A = \frac{\boldsymbol{n}^i . \boldsymbol{n}^j}{|\boldsymbol{n}^i| . |\boldsymbol{n}^j|}$$

where *i* and *j* are the reference numbers for 2 different plates. Based on the results in Table 1, we observe that the geometries are strongly based on highly symmetric polyhedrons, especially the 5 platonic solids. In fact, the PINNs solution for 6*p* lattice can be simply a dodecahedron, which cannot be tessellated cubically.[62] Note that even though the dodecahedron has 12 faces, its face normals only occupy 6 unique axes and hence, qualifies as a 6*p* structure. Similarly, the solution for 7*p* lattice indicates a mix of Cuboctahedron, Octahedron/ Tetrahedron and Cube, which actually returns the *p*SC-*p*FCC structure that is already known to exhibit isotropy and stiffness at the H-S upper limit[36,37]. This result not only lends credence to the validity of our analysis and the PINNs solutions, but importantly, highlights the difficulty of deriving ultrastiff, isotropic geometries based on heuristics – if the hybridization of the Plate Octet and Plate Simple Cubic structures had not produced the necessary mix of Cuboctahedron, Octahedron/ Tetrahedron and Cube geometries, the first isotropic designs with H-S stiffness[36,37] may not have been discovered. This may also be why 6*p* and 7*p* – 10*p* structures have yet to be reported in the literature.

As more plates become available to construct the isotropic PINNs lattices, the solutions indicate that the hybridization of more polyhedrons (*e.g.* rhombic dodecahedron), especially those with more faces (*e.g.* icosahedron), is needed. Overall, however, the geometries would resemble these polyhedrons less, as there is an increasing number of dihedral angles that do not match those from these polyhedrons.

In general, attaining isotropy becomes more straightforward as the number of plates increases, since more loading axes can be addressed[63]. However, a smaller number of plates is often desired as a matter of practicality, since simpler designs reduce both modelling complications, as well as manufacturing difficulties. Therefore, considering that previous isotropic lattice designs required a minimum of 7 plates (*p*SC-*p*FCC *a.k.a.* Plate Cubic Octet[36]), the present discovery that isotropy can be attained with only 6 plates represents a significant breakthrough.

Table 1: Dihedral angles between different plates for each *Xp* PINNs lattice geometry. A cell is highlighted when its angle falls within ± 5% of the dihedral angle of a known polyhedron. Note that internal angles of the polyhedrons here are typically obtuse, but a negative face normal vector, which does not affect the plate orientation, could lead to an acute angle instead, and hence, both are included in the legend.

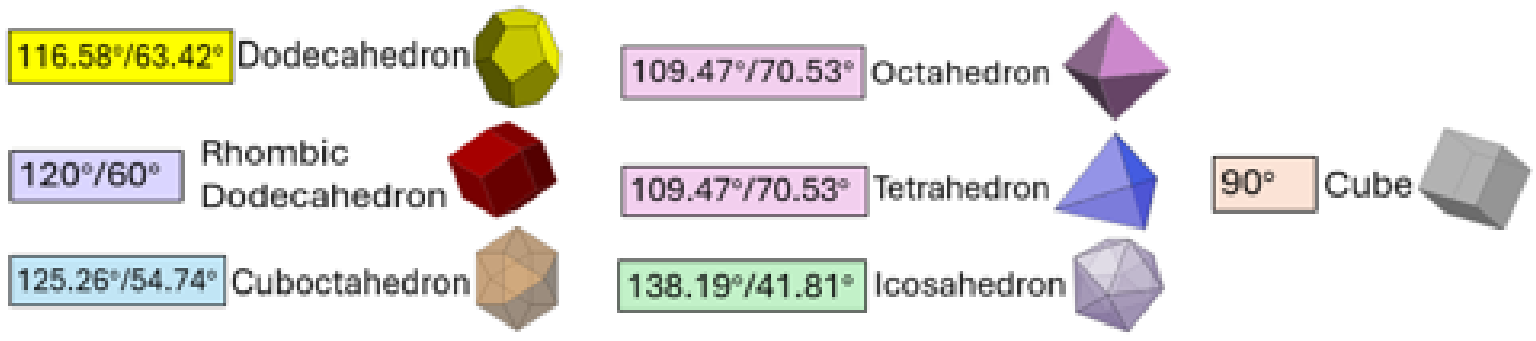

| *6p* | | *i* 1 | 2 | 3 | 4 | 5 |
|---|---|---|---|---|---|---|
| *j* | 2 | 63.42° | | | | |
| | 3 | 116.55° | 63.50° | | | |
| | 4 | 116.54° | 116.59° | 63.43° | | |
| | 5 | 63.44° | 63.48° | 63.43° | 63.40° | |
| | 6 | 63.46° | 116.57° | 116.56° | 63.47° | 63.45° |

| *7p* | | *i* 1 | 2 | 3 | 4 | 5 | 6 |
|---|---|---|---|---|---|---|---|
| *j* | 2 | 124.53° | | | | | |
| | 3 | 90.41° | 53.69° | | | | |
| | 4 | 125.64° | 71.02° | 124.61° | | | |
| | 5 | 89.93° | 124.75° | 89.50° | 125.28° | | |
| | 6 | 55.59° | 109.07° | 125.52° | 70.55° | 126.22° | |
| | 7 | 55.39° | 69.55° | 54.95° | 109.48° | 125.17° | 70.65° |

| *8p* | | *i* 1 | 2 | 3 | 4 | 5 | 6 | 7 |
|---|---|---|---|---|---|---|---|---|
| *j* | 2 | 116.65° | | | | | | |
| | 3 | 124.17° | 49.90° | | | | | |
| | 4 | 108.26° | 127.87° | 120.09° | | | | |
| | 5 | 63.35° | 135.85° | 91.92° | 86.41° | | | |
| | 6 | 128.75° | 110.58° | 74.56° | 48.94° | 69.42° | | |
| | 7 | 124.17° | 88.12° | 110.66° | 43.56° | 130.04° | 74.56° | |
| | 8 | 108.26° | 93.55° | 43.56° | 96.11° | 52.18° | 48.94° | 120.09° |

| *9p* | | *i* 1 | 2 | 3 | 4 | 5 | 6 | 7 | 8 |
|---|---|---|---|---|---|---|---|---|---|
| *j* | 2 | 94.43° | | | | | | | |
| | 3 | 113.15° | 99.69° | | | | | | |
| | 4 | 139.89° | 49.55° | 65.67° | | | | | |
| | 5 | 70.42° | 134.91° | 54.05° | 119.62° | | | | |
| | 6 | 48.72° | 53.49° | 136.87° | 102.62° | 117.89° | | | |
| | 7 | 128.29° | 84.64° | 117.94° | 72.47° | 138.41° | 94.68° | | |
| | 8 | 57.71° | 49.27° | 85.38° | 82.81° | 89.14° | 51.55° | 132.41° | |
| | 9 | 79.15° | 120.47° | 137.51° | 130.76° | 98.70° | 82.72° | 58.26° | 130.51° |

| *10p* | | *i* 1 | 2 | 3 | 4 | 5 | 6 | 7 | 8 | 9 |
|---|---|---|---|---|---|---|---|---|---|---|
| *j* | 2 | 140.12° | | | | | | | | |
| | 3 | 101.41° | 58.02° | | | | | | | |
| | 4 | 112.29° | 63.46° | 118.66° | | | | | | |
| | 5 | 107.78° | 97.40° | 71.10° | 134.83° | | | | | |
| | 6 | 43.31° | 97.36° | 81.42° | 89.64° | 134.84° | | | | |
| | 7 | 105.95° | 37.57° | 69.72° | 52.80° | 132.15° | 63.08° | | | |
| | 8 | 48.85° | 110.56° | 55.27° | 145.85° | 77.97° | 56.94° | 100.69° | | |
| | 9 | 51.89° | 123.08° | 142.95° | 64.31° | 136.06° | 61.54° | 91.65° | 100.16° | |
| | 10 | 77.37° | 65.19° | 40.73° | 98.90° | 109.54° | 42.92° | 47.85° | 65.81° | 103.08° |

## 4.2 Ultrastiff Isotropic Lattice Designs

Even though the plate orientations in the lattices are known from Eq. (12) – (16), the solutions provide no information on the optimal placements of the plates. While it is tempting to simply follow closely to the polyhedron geometries, the structures obtained this way may not be able to be tessellated into lattices, as is the case with the dodecahedron design for *6p* lattices. Furthermore, the polyhedron geometries themselves may not provide sufficient clues for a deterministic placement of all the plates. Therefore, to eliminate such uncertainty, we developed a simple, standardized procedure for the lattice construction.

Firstly, the requisite number of plates were oriented in cubic space according to the normal vector solutions provided in Eq. (12) – (16). Next, the plates were positioned such that their centers of mass coincided at a common spatial point. Boolean union was then applied to join all the plates into a single part. Figure 2A shows the resultant structures for $X = 6, 7, 8, 9$ and 10 plates. Note that each plate had been assigned the same volume in our analysis, but because their areas were different, the thickness varied from plate to plate.

In a departure from previous isotropic lattice designs[36,37,51,52,54,57], it can be observed that these PINNs lattices lack cubic symmetry, resembling stochastic foam structures[36,64] more than periodic hybrid unit cells. As a result, the structures in Figure 2A cannot be tessellated into cubic lattices and had to be added to their mirror images in the *X*, *Y*, and *Z* directions to form a periodic unit cell first (Figure 2B). Finite element simulations were then conducted on 2 × 2 × 2 of the subunit cells (or mirrored periodic unit cell) to obtain their stress distributions under compressive loading (Figure 2C) and shear loading (Figure 2D), as well as the lattice properties (Figure 3).

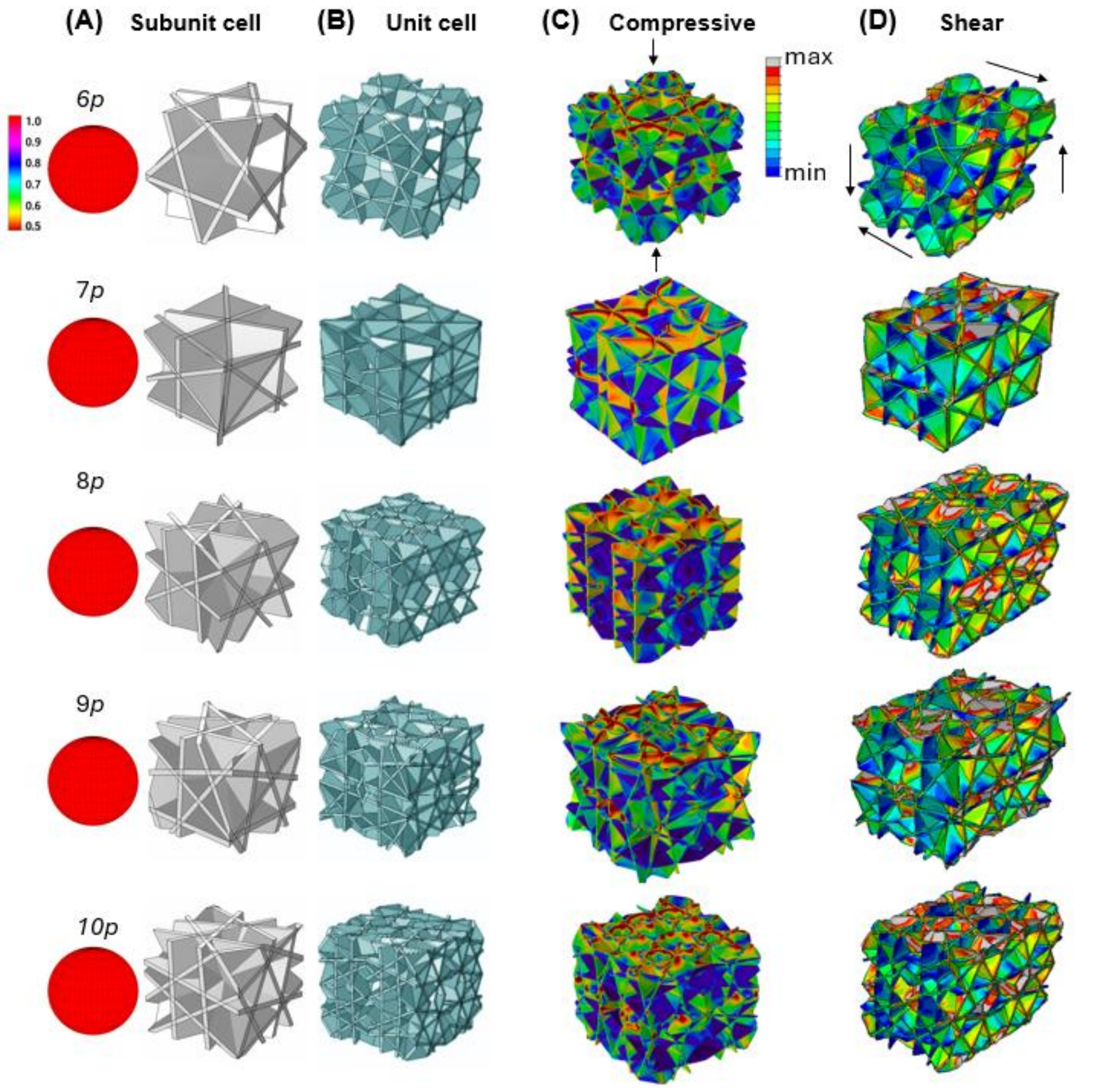

**Figure 2: (A)** Subunit cells, **(B)** Unit cells, **(C)** Compressive loading, and **(D)** shear loading of the *Xp* PINNs lattices. The white colored subunit cells have non-symmetric faces and cannot be tessellated, while the corresponding green colored lattices can be tessellated to make a 3D structure, and treated as unit cells of the respective lattices namely *6p, 7p, 8p, 9p* and *10p lattices*. The 3D polar plot (leftmost) shows the theoretical directional stiffness of the corresponding subunit cells.

### 4.3 Validation

Under quasistatic loading, stresses in the *Xp* PINNs lattices are observed to concentrate in plates that are most closely aligned to the loading axes - <100> for compression (Figure 2C) and <110> for shear loading (Figure 2D), which is expected. Compression loading was used to determine the effective stiffness ($\bar{E}$) and Poisson's ratio ($\bar{\nu}$), whereas shear loading was used to evaluate the effective shear modulus ( $\bar{G}$ ). These effective elastic properties were subsequently utilized to calculate the Zener ratio ($Z=2\bar{G}(1+\bar{\nu})/\bar{E}$), which serves as a measure of elastic anisotropy. However, significant bending stresses were also found in the kinks of the plates (Figure 2C and 2D), which were formed when inclined plates in the subunit cells (Figure 2A) joined with their mirror images at an angle to produce the periodic unit cell (Figure 2B). Under load, these kinks generate moments that would lead to bending/ buckling of the plates, which violates a key assumption of our analysis – that deformations would only be in-plane.

Because of this, the isotropic PINNs lattice moduli were compromised and their simulated relative stiffness, $\bar{E}/E_s$, could only attain an average of 84.2 ± 6 % of the H-S upper bound for $0.1 \leq \rho_R \leq 0.5$ (Figure 3C). Similarly, the relative shear modulus, $\bar{G}/G_s$ (Figure 3D), and Poisson's ratio, $\bar{\nu}$ (Figure 3E), of the lattices are close to, but consistently fall short of the H–S upper bound. In addition, the simulated $\bar{G}/G_s$ values are observed to diverge from the H-S upper bound with increasing relative density, partly due to the breakdown of the plane stress assumption as the plates became thicker. Since the Zener ratio is given by $Z = 2\bar{G}(1+\bar{\nu})/\bar{E}$, it is no surprise that the isotropy of the PINNs dovetails the trend for relative shear modulus, starting at ~ 1 (isotropic) for low relative density and gradually falling (*i.e.* becoming more anisotropic) towards an asymptotic value of ~ 0.8 as relative density increases (Figure 3F).

To further confirm that plate positioning affects the stiffness and isotropy of these lattices, we rearranged the plates in the *pSC-pFCC* design[36,37] according to our methodology outlined in Section 4.2 *i.e.*, position all the plates so their centres of mass coincide to obtain a unit cell. Finite element simulations show that the relative stiffness, relative shear modulus and Zener ratio were lower for the rearranged geometry compared to the original *pSC-pFCC* design, clearly indicating that plate positions should be optimized to achieve the best results.

Interestingly, the mechanical properties of the rearranged *pSC-pFCC* geometry are similar to those of the *Xp* lattices. In the same way, we can also observe from Figure 3C – 3F that there is not much differentiation in the stiffness, shear modulus and anisotropy amongst the *Xp* PINNs lattices (along with designed rearranged *pSC-FCC* lattice), except for Poisson's ratio, but even that occurs over a very narrow range of 0.16 – 0.23. This result indicates that the optimization of plate positions is a stronger determinant of elastic properties than the number of plates in the subunit cell of a lattice. Therefore, solving for geometries with more than 10 plates would likely not yield better results or more insights.

Experimental validation of the finite element simulation results was carried out with additively manufactured stainless steel lattices of the 6*p* PINNs design ($\rho_R = 0.45$) (Figure 3C). This geometry was selected because it is the first to suggest that isotropic lattices can be constructed with less than 7 plates, the minimum currently required in the literature[36,37]. From Figures 3A and 3B, it can be observed that the 6*p* PINNs geometry was faithfully reproduced with the LAPIS technique. Micro-CT examination revealed that the internal voids were indeed empty as intended. Analysis of their elastic responses indicates a stiffness within ± 10% of the simulated values for both the <100> and <110> orientations (Figure 3C; Table 2). These experimental values are also in the vicinity of the H-S upper bound, falling between 86% to 100% of the theoretical limit. From the experimental results (Table 2), the ratio of specific relative modulus - <110> to <100> is 0.89, which is well-aligned with the simulated value of 0.84. Such good agreement between the analytical, simulated and experimental data, for both <100> and <110> lend significant confidence to our solutions for isotropic plate lattices with ultrahigh stiffnesses.

Table 2: Specific relative stiffness, $\bar{E}/E_s\rho_R$, of isotropic 6*p* PINNs lattices in different orientations.

| | Analytical (Hashin-Shtrikman Upper Bound) | Simulation | Experimental |
|---|---|---|---|
| 6*p* <100> | 0.63 | 0.57 | 0.63 ± 0.077 |
| 6*p* <110> | 0.65 | 0.60 | 0.56 ± 0.053 |

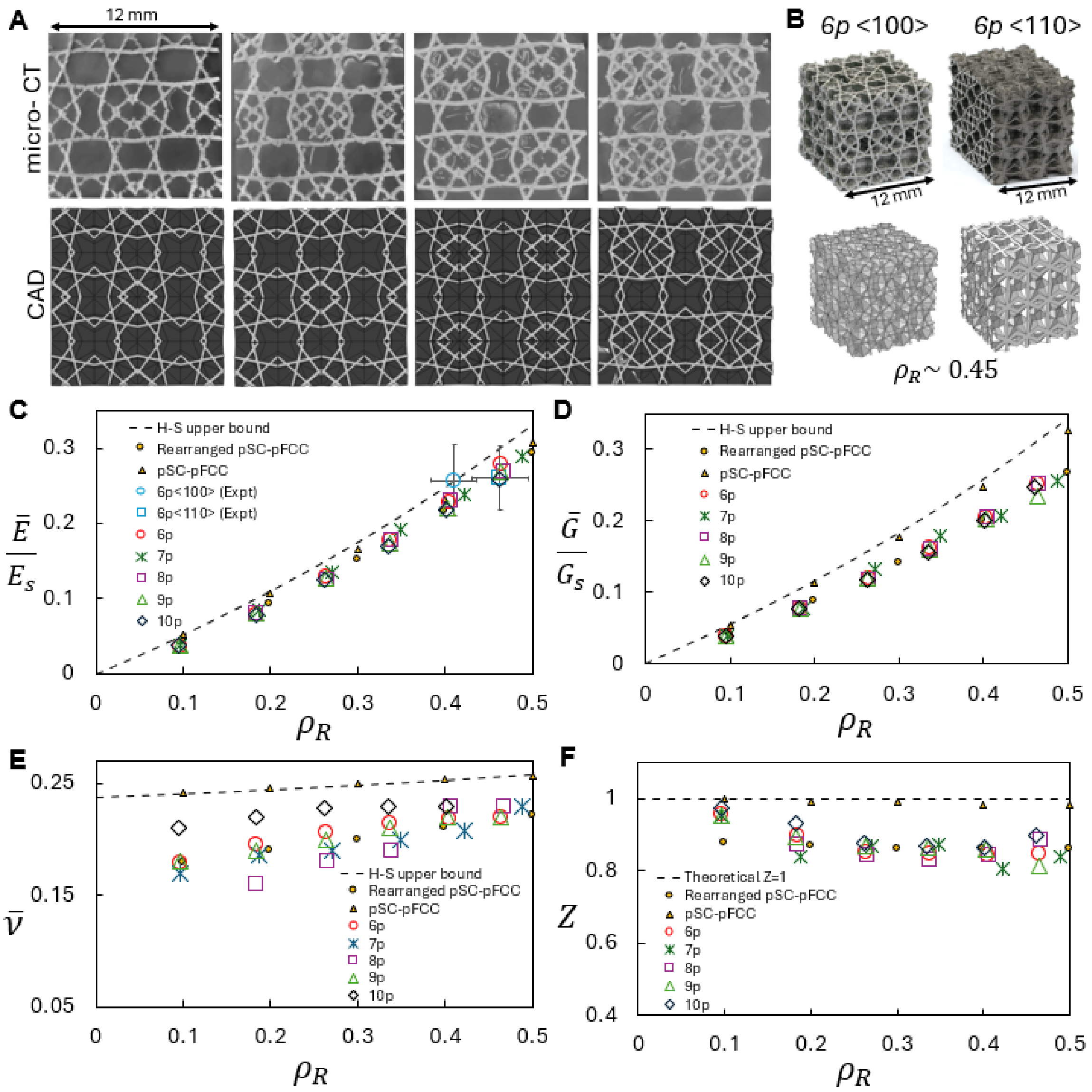

**Figure 3: (A)** Micro-CT scan images of fabricated 6*p* lattice along with cross-sections of corresponding CAD model, **(B)** Fabricated specimens of 6*p* <100> and 6*p* <110> along with corresponding CAD model, Effective **(C)** relative stiffness ($\bar{E}/E_s$), **(D)** relative shear modulus ($\bar{G}/G_s$), **(E)** Poisson's ratio, and **(F)** Zener ratio ($Z$=$2\bar{G}(1+\bar{\nu})/\bar{E}$), with respect to relative density ($\rho_R$) of *Xp* lattices. The rearranged *pSC-pFCC* lattice is designed as discussed in Section 4.2 and FE results are plotted for comparison. The Hashin-Shtrikman upper bound (black dashed line) has been plotted as a reference.

## 4.4 Benchmarking

Comparing the specific relative stiffness, $\bar{E}/E_s\rho_R$, of the PINNs lattices to those of previously reported structures, *pSC-pFCC* (Plate SC-FCC)[36,37,51], Plate SC-BCC[37] and compound nested lattices (1*pSC*:216*pBCC* and 1*pSC*:512*pFCC*)[52], it can be observed that the PINNs lattices are softer at low $\rho_R$, due to the aforementioned bending of kinked plates (Figure 4). Moreover, the comparison between *pSC-pFCC* and *7p* PINNs lattices, which have the same number of plates and plate orientations, shows that judicious placement of plates can significantly improve the mechanical performance of the lattice, in line with observations from Figure 3.

However, as the relative density increases, the stiffness of most structures deviates away from the H-S upper bound, while that of the PINNs lattices advance closer, eventually converging with that of *pSC-pFCC* at $\rho_R \sim 0.45$. For instance, for the *6p* PINNs lattice, its relative specific stiffness can be observed to increase from 73% of H-S upper bound at $\rho_R$ = 0.1, to 90% of H-S upper bound at $\rho_R$ = 0.4 (Figure 4). The trend is similarly observed for the relative shear modulus ($\bar{G}/G_s$) and Poisson's ratio ($\bar{\nu}$) across all *Xp* PINNs lattices (Figure 3D – 3E). This is because, as the plates thicken at higher relative densities, the buckling of the plates at the kinks becomes less severe and the designs are able to better realize the same load-bearing potential they have with *pSC-pFCC*. As a result, for $\rho_R \geq 0.1$, the *Xp* PINNs lattices outperform most of the isotropic designs reported previously, such as isotropic truss[32], hierarchical isotropic truss plate hybrid structure[53], topology-optimized structures[54], as well as closed cell foams - quasi-random[36], Voronoi and Gaussian random tessellated[64], Kelvin-BCC and bi-disperse FCC foams[65] - which they morphologically resemble the most.

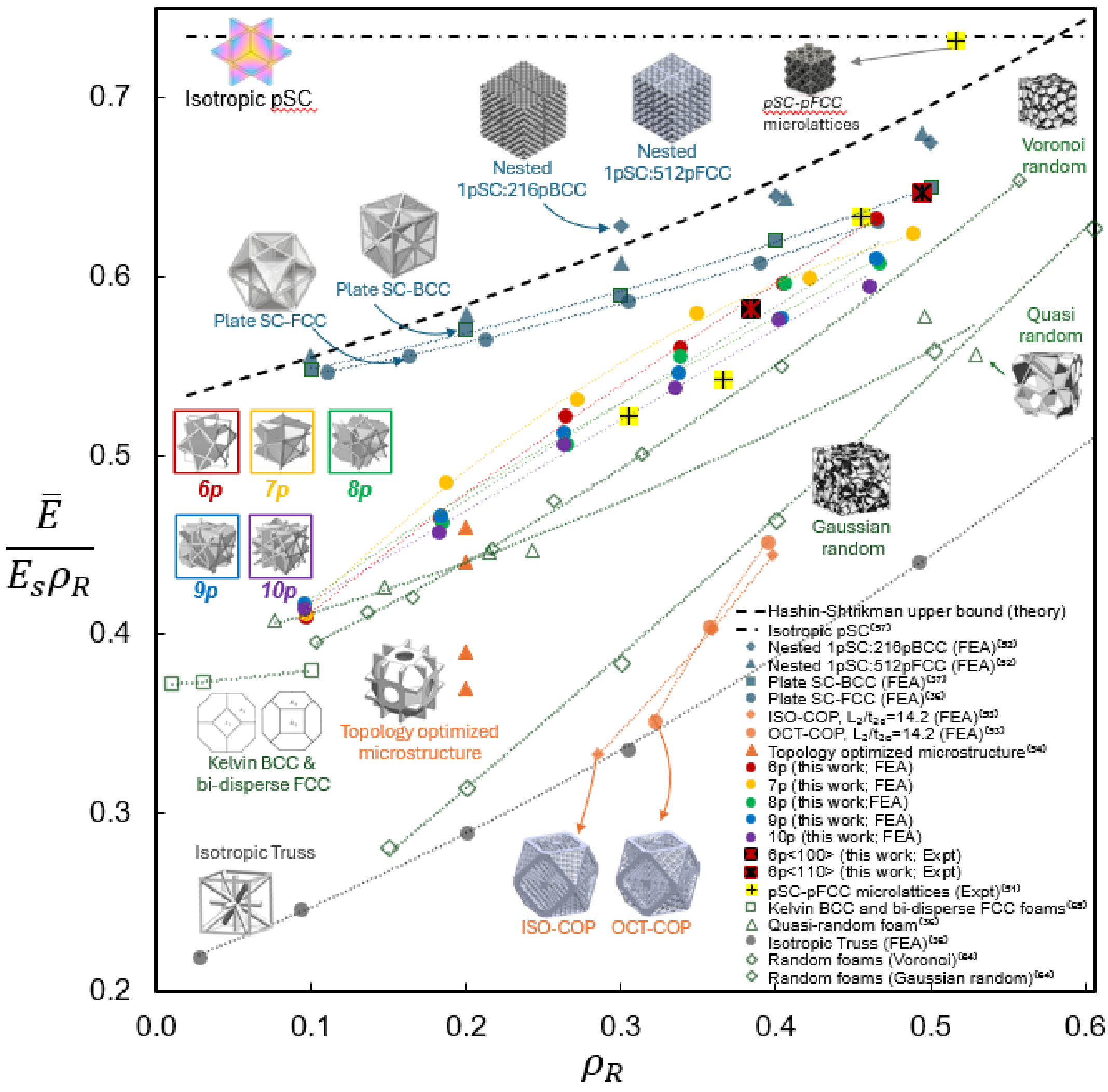

**Figure 4:** Comparison of specific relative modulus for different lattice structures[36,51–54,57,64,65]. The isotropic plate Simple Cubic (Isotropic pSC) made from anisotropic material is included for benchmarking.

On the other hand, the Zener ratio ($Z$) of the *Xp* PINNs lattices initially shows strong isotropy ($Z$ = 0.95) for $\rho_R$ = 0.1, but registers a modest deviation at higher relative densities (Figure 5). Compared to other isotropic geometries, such as isotropic truss[36], quasi-random foams[36], *p*SC-*p*FCC (Plate SC-FCC) lattices[36] and Plate SC-BCC[37], the *Xp* lattices are not as isotropic at higher relative densities. Nevertheless, their Zener ratios remain closer to unity than highly anisotropic designs like the plate Octet and plate Simple Cubic (*p*SC) lattices.

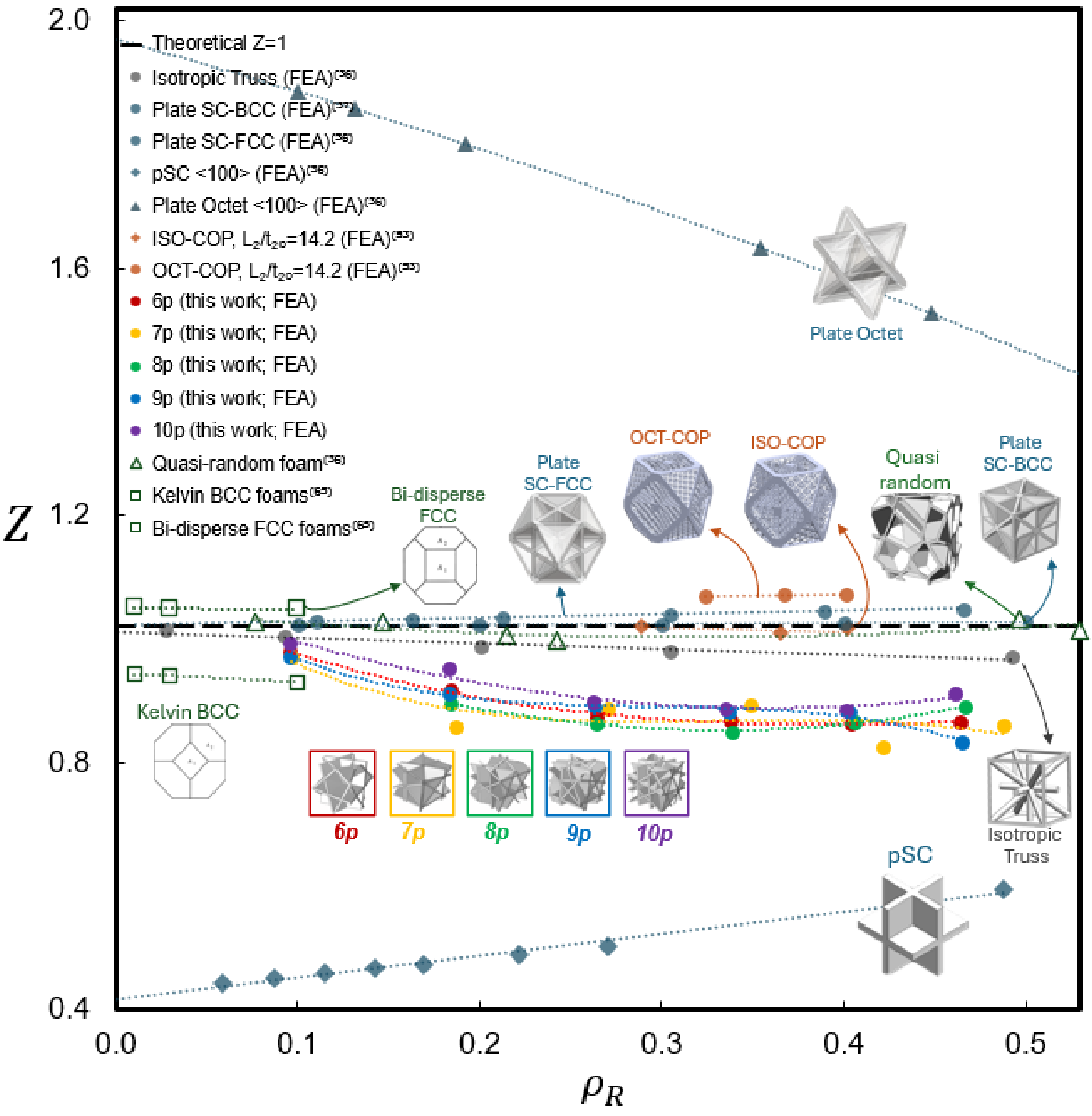

**Figure 5:** Comparison of Zener ratio (Z) for different lattice structures[36,37,53,65]. The plate Octet <100> and plate Simple Cubic (*p*SC) <100> lattices are anisotropic and have been included for reference only.

## 5 Discussion

Unlike conventional topology optimization, heuristic or evolutionary approaches that search for desirable geometries through iterative exploration of predefined design spaces, the present framework formulates lattice design as a physics-constrained problem. By treating plate orientations as unknowns and directly enforcing the isotropy conditions through Physics-Informed Neural Networks, we show that previously unknown isotropic plate lattices can be discovered systematically without relying on geometric intuition, symmetry assumptions, or large-scale parametric optimization. This enables the identification of novel plate arrangements with near-optimal stiffness that would be difficult to obtain using traditional design methodologies.

Interestingly, the plate lattices obtained through the proposed PINNs framework exhibit geometric characteristics similar to stochastic cellular solids (quasi-random foams), despite being derived from a completely deterministic design procedure. This similarity arises because both systems seek a nearly uniform distribution of load-carrying elements in orientation space, a fundamental requirement for isotropic mechanical behaviour. Unlike quasi-random foams, however, the proposed architectures satisfy the isotropy conditions analytically and reproducibly, eliminating the need for randomness while retaining many of the desirable characteristics of foam-like microstructures such as quasi-isotropy and high stiffness. The *Xp* PINNs lattices may, therefore, be interpreted as deterministic minimal-complexity analogues of isotropic foams, albeit with better specific stiffness.

While previously reported hybrid architectures such as *p*SC–*p*FCC and nested 1*p*SC:512*p*FCC lattices achieve higher stiffness and approach the Hashin–Shtrikman upper bound more closely, they rely on highly symmetric and geometrically complex topologies derived through heuristic design strategies. In contrast, the present lattices emerge directly from the solution of isotropy equations without imposing any predefined crystal symmetry or lattice motif. The resulting architectures can achieve isotropy with as few as six plates resembling dodecahedron, representing a near-minimal isotropic realization. Although their stiffness is somewhat lower than that of optimized hybrid lattices because this method does not solve for the position of the plates in the lattices for optimized stiffness, they offer important advantages in terms of systematic inverse design. Therefore, the proposed lattices establish a new class of deterministic isotropic metamaterials that complement, rather than compete directly with, highly optimized hybrid architectures.

Beyond isotropic metamaterials made from the isotropic constituent material, the same design philosophy can potentially be extended to isotropic metamaterial made from anisotropic crystals or composite materials, where the network would solve directly for microstructural configurations to achieve prescribed effective properties. Consequently, the broader significance of the work lies not only in the specific isotropic lattices obtained, but also in the establishment of a general physics-constrained machine-learning framework for the systematic design of multifunctional architected materials.

## 6 Conclusions

In this study, novel lattice structures are obtained using PINNs by satisfying the isotropy conditions of plate lattices. The FE and experimental results demonstrated agreement with the analytically derived effective stiffness values and Zener ratio, thereby validating both isotropic design and the accuracy of the solution. The obtained PINNs results for *6p* and *7p* lattices returned the dodecahedron and earlier reported *pSC-pFCC* design respectively, which were known to be isotropic and, thus, provided evidence for the veracity of the results. However, non-optimized placement of the plates resulted in the formation of kinked plates that bend under load, causing the stiffness and isotropy to deviate from the expected values. Nevertheless, the present advancement contributes to the broader field of mechanical metamaterials by providing a new pathway for developing lightweight, high-performance materials suitable for demanding structural applications in aerospace, automotive, biomedical, and civil engineering domains.

## Supplementary Information

Additional details of the finite element simulations and results can be found in the Supplementary Material.

## Acknowledgements

The authors declare no conflicts of interest. This work was supported by C.Q.L's Nanyang Assistant Professorship Grant (award no.: #022081–00001) and A*Star RIE2025 MTC Strategic & Translational Research Fund (M25N2K0101).

## Data Availability

The data that supports the findings of this study are available within the article and its supplementary material.

# Supplementary Information

## Derivation of Novel Isotropic Metamaterials at the Hashin-Shtrikman Stiffness Limit Enabled by Physics Informed Neural Networks

Manish Kumar Singh[1], Dominic Kang Jueh Lim[1,2], Anson Wen Han Cheong[1,2], Adams Wai Kin Kong[3], Chang Quan Lai[1,2,4*],

[1]*School of Mechanical and Aerospace Engineering, Nanyang Technological University, 50 Nanyang Ave, Singapore, 639798, Singapore*

[2]*Singapore Centre for 3D Printing, 63 Nanyang Drive, Singapore 636922, Singapore*

[3]*College of Computing and Data Science, Nanyang Technological University, 50 Nanyang Ave, Singapore, 639798, Singapore*

[4]*School of Materials Science & Engineering, Nanyang Technological University, 50 Nanyang Ave, Singapore, 639798, Singapore*

[*] cqlai@ntu.edu.sg

# 7 Application of the Abaqus plugin, EasyPBC

The EasyPBC plugin tool for ABAQUS[1] is employed to evaluate the homogenized elastic properties of the proposed lattice structure. The plugin enables the implementation of periodic boundary conditions (PBCs) and the determination of homogenized elastic properties from a representative volume element (RVE) of a composite material. Unlike conventional composite materials, where the RVE consists of multiple constituent phases (e.g., matrix and reinforcement), the present lattice structures are cellular architectures containing large void regions. To implement the PBCs using EasyPBC plugin, the unit cell had to be embedded within a cube that consisted of a material with negligible mechanical stiffness ($10^{-6}$) and Poisson's ratio ($10^{-6}$), effectively mimicking a void. In contrast, the lattice material was assigned standard values of 1 MPa for Young's modulus ($E_s$) and 0.3 for Poisson's ratio ($\nu_s$) to facilitate dimensionless analysis used for benchmarking against analytical predictions and other lattice configurations.

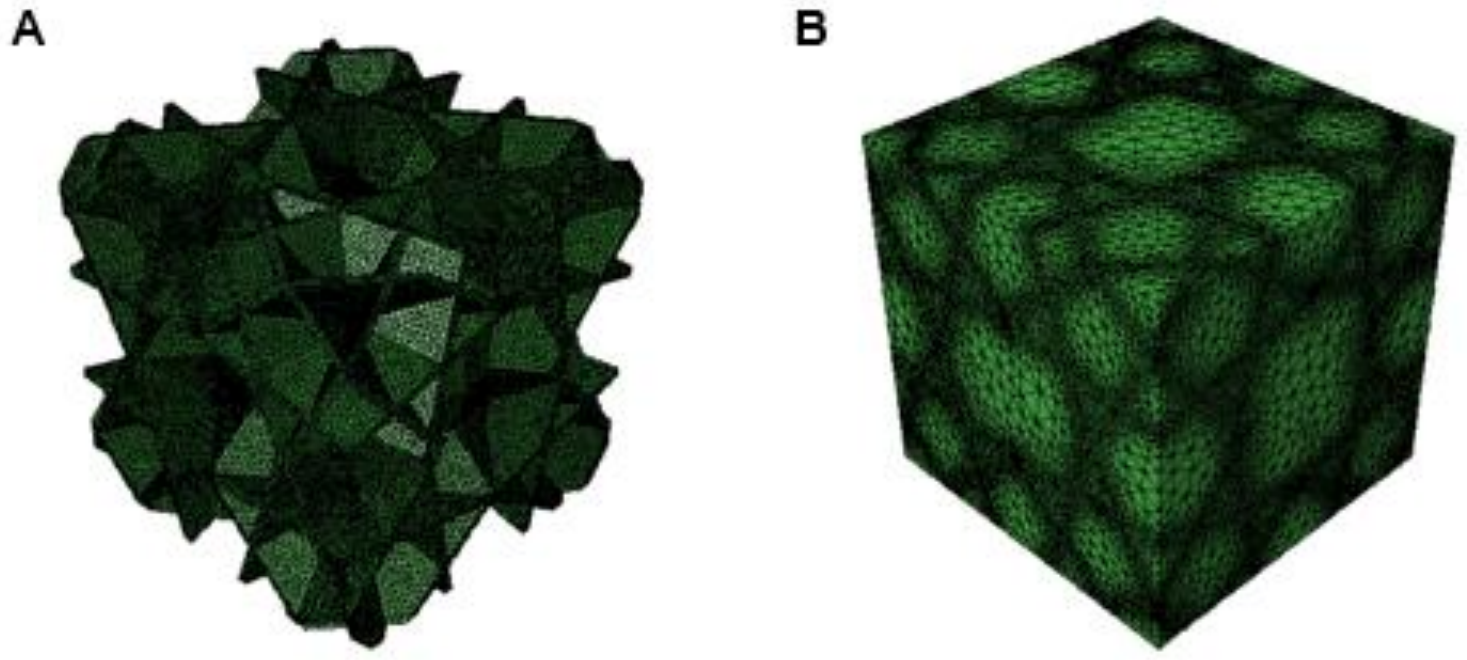

**Figure S1:** **(A)** 3D mesh structure of *6p* unit cell and **(B)** the unit cell embedded within a surrounding cubic domain composed of a material with negligible mechanical stiffness, effectively mimicking a void. This auxiliary domain (surrounding cube) provided a continuous geometry required for the application of EasyPBC plugin while contributing negligibly to the overall mechanical response of the structure.

# 8 Finite Element Convergence Analysis

To ensure the accuracy and reliability of the finite element analysis (FEA), a mesh convergence study was conducted for the *6p* lattice structure at a relative density of 0.3. The study evaluates the sensitivity of the computed effective elastic properties to mesh refinement using 3D 10-node quadratic tetrahedral elements (C3D10M).

Specifically, the following dimensionless homogenized properties were analyzed as functions of mesh density: Effective relative modulus ($\bar{E}/E_S$), effective normalized shear modulus ($\bar{G}/E_S$) and effective Poisson's ratio ($\bar{\nu}$). Figure S3 presents the resulting trends, clearly indicating that the effective moduli stabilize beyond a certain mesh density, thereby confirming numerical convergence.

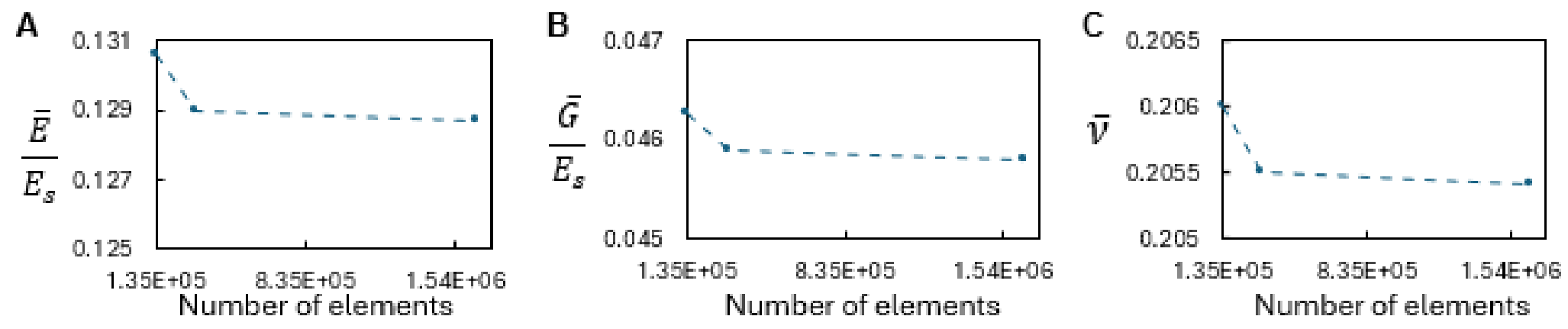

**Supplementary Figure S2**: Mesh convergence plot for **(A)** relative Young's modulus, **(B)** normalized shear modulus, and **(C)** Poisson's ratio of *6p* unit cell.

# 9 Loss Function of PINNS

Figure S1 shows the mean squared error (MSE) loss evolution of the governing equations for the *5p, 6p, 7p, 8p, 9p*, and *10p* geometries. The MSE decreases rapidly during the initial 1000 epochs, followed by oscillations without significant improvement. Therefore, the training was terminated at 30,000 epochs. For the 5p lattice, the MSE is approximately $2\times10^{-3}$, which is insufficient to accurately satisfy the governing equations or achieve isotropy, also confirmed by finite element (FE) simulations. In contrast, lattices with more than five plates exhibit MSE values below $10^{-4}$, indicating that there are valid $X$ plate orientations that can satisfy the isotropy conditions within a small margin of error.

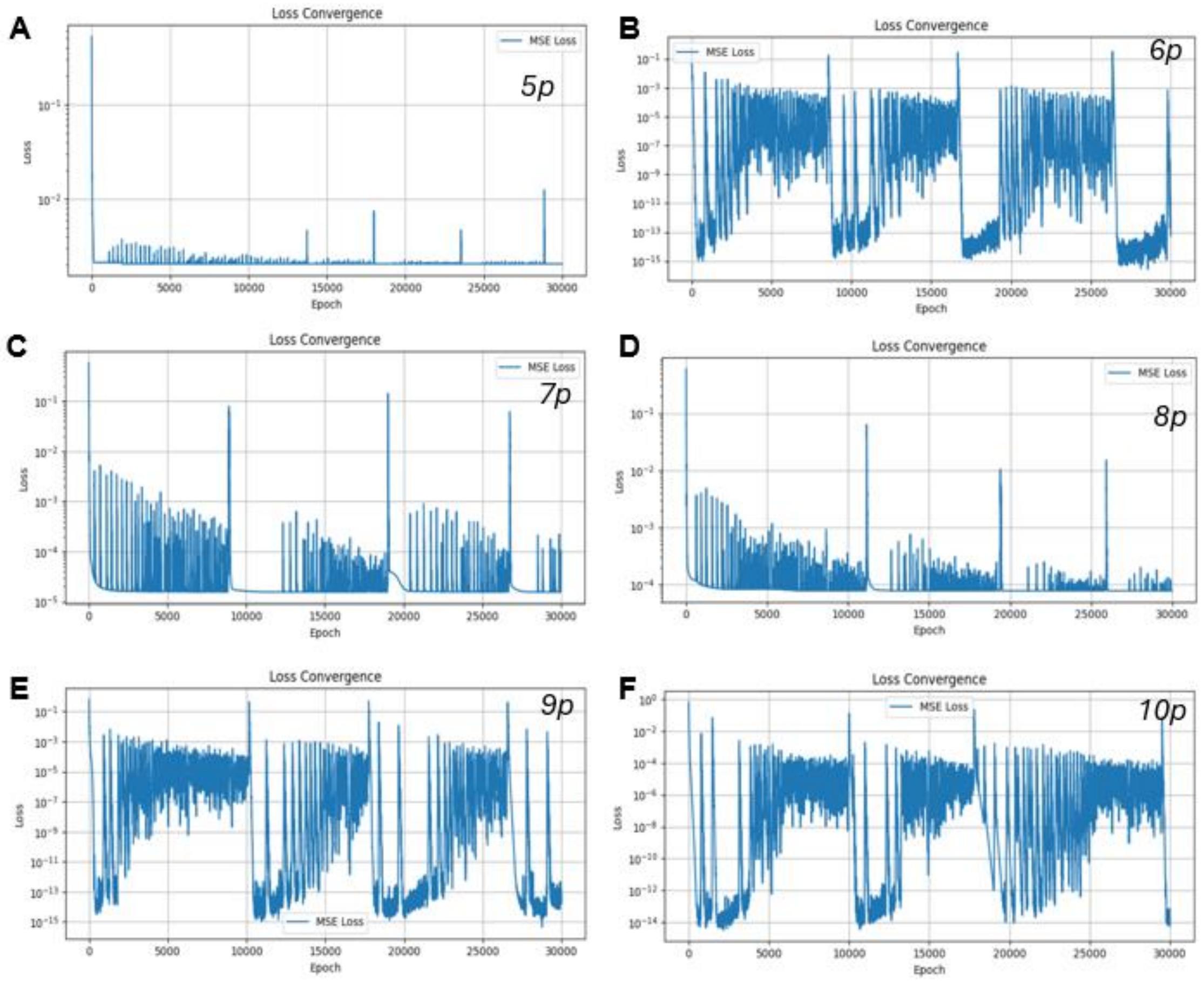

**Figure S3:** MSE loss with numbers of epoch in PINNs for **(A)** *5p*, **(B)** *6p*, **(C)** *7p*, **(D)** *8p*, **(E)** *9p* and **(F)** *10p* lattices.

## 10 Sample Code for 8p Lattices

```
# -*- coding: utf-8 -*-
"""Isotropic8p.ipynb

import torch
import torch.nn as nn
import torch.optim as optim
import matplotlib.pyplot as plt
import random
import numpy as np

# -------------------- Reproducibility --------------------
def set_seed(seed=42):
    random.seed(seed)
    np.random.seed(seed)
    torch.manual_seed(seed)
    torch.cuda.manual_seed_all(seed)
    torch.backends.cudnn.deterministic = True
    torch.backends.cudnn.benchmark = False

set_seed(42)  # Set seed before model/training

# -------------------- Define the PINN model --------------------
class PINN(nn.Module):
    def __init__(self, input_dim=24, hidden_dim=128, output_dim=24, num_layers=4):
        super(PINN, self).__init__()
        layers = [nn.Linear(input_dim, hidden_dim), nn.Tanh()]
        for _ in range(num_layers - 1):
            layers.append(nn.Linear(hidden_dim, hidden_dim))
            layers.append(nn.Tanh())
        layers.append(nn.Linear(hidden_dim, output_dim))
        self.network = nn.Sequential(*layers)

    def forward(self, x):
        return self.network(x)

# -------------------- Define the 24 constraints --------------------
def compute_constraints(output):
    n = output.view(8, 3)  # Reshape to (8, 3) for constraint definitions

    eqs = []

    # 8 unit norm constraints (||nᵢ||² = 1)
    for i in range(8):
        eqs.append(torch.sum(n[i] ** 2) - 1)

    # 3 second-order constraints
    eqs.append(3/8 * torch.sum(n[:, 0] ** 2) - 1)
    eqs.append(3/8 * torch.sum(n[:, 1] ** 2) - 1)
    eqs.append(3/8 * torch.sum(n[:, 2] ** 2) - 1)
```

```
    # 3 fourth-order constraints
    eqs.append(5/8 * torch.sum(n[:, 0] ** 4) - 1)
    eqs.append(5/8 * torch.sum(n[:, 1] ** 4) - 1)
    eqs.append(5/8 * torch.sum(n[:, 2] ** 4) - 1)

    # 9 mixed moment constraints
    eqs.append(torch.sum(n[:, 0]**3 * n[:, 1]))
    eqs.append(torch.sum(n[:, 0]**3 * n[:, 2]))
    eqs.append(torch.sum(n[:, 0]**2 * n[:, 1] * n[:, 2]))
    eqs.append(torch.sum(n[:, 1]**3 * n[:, 0]))
    eqs.append(torch.sum(n[:, 1]**2 * n[:, 0] * n[:, 2]))
    eqs.append(torch.sum(n[:, 1]**3 * n[:, 2]))
    eqs.append(torch.sum(n[:, 2]**3 * n[:, 0]))
    eqs.append(torch.sum(n[:, 2]**3 * n[:, 1]))
    eqs.append(torch.sum(n[:, 2]**2 * n[:, 0] * n[:, 1]))

    return torch.stack(eqs)  # 23 constraints

# -------------------- Training function --------------------
def train_PINN(model, epochs=30000, lr=0.001):
    optimizer = optim.Adam(model.parameters(), lr=lr)
    criterion = nn.MSELoss()
    loss_history = []

    for epoch in range(epochs):
        optimizer.zero_grad()

        x_input = torch.ones((1, 24))  # Fixed input vector
        output = model(x_input)

        constraints = compute_constraints(output)
        loss = criterion(constraints, torch.zeros_like(constraints))

        loss.backward()
        optimizer.step()
        loss_history.append(loss.item())

        if epoch % 1000 == 0:
            print(f"Epoch {epoch}, Loss: {loss.item():.8f}")

    return model, loss_history

# -------------------- Instantiate and train --------------------
model = PINN()
trained_model, loss_history = train_PINN(model)

# -------------------- Final prediction --------------------
with torch.no_grad():
    x_input = torch.ones((1, 24))
    predicted = trained_model(x_input).numpy()
```

```python
print("\nPredicted vector (flattened):")
print(predicted)

# -------------------- Plotting --------------------
plt.figure(figsize=(8, 4))
plt.plot(loss_history, label="MSE Loss")
plt.yscale("log")
plt.xlabel("Epoch")
plt.ylabel("Loss")
plt.title("Loss Convergence")
plt.grid(True)
plt.legend()
plt.tight_layout()
plt.show()
```